\documentclass[usenatbib]{mnras}
\usepackage{graphicx}

\title[New HH objects around BBW 192E]{Herbig-Haro flows around BBWo 192E (GM 1-23) nebula}

\author[T.Yu. Magakian et al.]{T.Yu. Magakian$^{1}$\thanks{E-mail:tigmag@sci.am}, T.A. Movsessian$^{1}$, H.R. Andreasyan$^{1}$, J.Bally$^{2}$ and A.S. Rastorguev$^{3,4}$\\
$^1$Byurakan Observatory of the Nat.Acad.Sci. of Armenia, Byurakan, Aragatsotn prov., 0213, Armenia\\
$^2$Department of Astrophysical and Planetary Sciences, University of Colorado, Boulder, CO,USA\\
$^3$Faculty of Physics, Lomonosov Moscow State University, Leninskie Gory 1, bldg.2, Moscow, Russia\\
$^4$Sternberg Astronomical Institute, Lomonosov Moscow State University, Universitetskii prospect 13, Moscow, Russia
}%

\begin{document}

\date{}

\pagerange{\pageref{firstpage}--\pageref{lastpage}} \pubyear{2020}

\maketitle
\label{firstpage}

\begin{abstract}
  
We studied a small comet-shape reflection nebula, located in the dark cloud SL~4 in the Vela Molecular Ridge cloud C, known as BBWo~192E (GM~1-23), and a young infrared cluster embedded into the nebula, for the evidences of recent star formation.
We obtained the images of BBWo~192E in H$\alpha$ and  [S~\textsc{ii}] lines
and in SDSS i' with  Blanco telescope at the Cerro Tololo Interamerican
Observatory to discover new Herbig-Haro (HH) flows. 2MASS and WISE  surveys were used for the search of the additional member stars of the cluster.  We also studied proper motions and parallaxes of the cluster members with the aid of GAIA DR2.   
Five new groups containing at least 9 HH objects tracing several distinct outflows were revealed. A previously unreported reflection nebula and a number of probable outflow sources were found in the infrared range. The proper motions allowed selecting eight probable member stars in the visual range. Their parallaxes correspond to a mean distance $800 \pm 100$ pc for this cluster. The bolometric luminosities  of the brightest cluster members are 1010 $L_{\odot}$ (IRAS 08513$-$4201, the strong source in the center of the cluster) and   2 to  6 $L_{\odot}$  for the five other stars. The  existence of the optical HH flows around   the    infrared cluster of YSOs suggests that star formation in this cloud is on-going around the  more massive HAeBe star.  By its morphology and  other features this  star-forming region is similar to the zone of star formation near CPM~19.  
\end{abstract}


\section{Introduction}
\label{sec:intro}

A small comet-shape reflection nebula, located in the elongated dark cloud SL 4 \citep{SL} in the Vela Molecular Ridge cloud C \citep{MM},  is known as GM~1-23 \citep{GM} and BBWo~192E \citep{BBW}. Such nebulous objects are often indicators of ongoing star formation and this case is no exception. The survey of \citet{PR} revealed tens of H$\alpha$ emission-line stars in the vicinity of SL~4, but did not find any Herbig-Haro (HH) objects.    The BBWo~192E nebula  was studied for the first time in this  work;  the authors concluded that it was a reflection nebula and that the relatively bright star on its northern edge is a projected foreground object. They suggested that the illuminating star is embedded in the cloud and should be HAeBe type young star.

This BBWo~192E nebula is associated with a bright and very red source,  IRAS 08513--4201, recognized as a Class I object with a near-IR counterpart \citep{Liseau}.   Further multi-frequency studies \citep{Burkert,Massi,Dutra} in the near and mid-infrared revealed a young infrared cluster embedded in the nebula.   IRAS 08513--4201 was found to be its most luminous member, illuminating not only the optical nebula but also a bipolar IR reflection nebula.   The distance of this group was estimated by \citet{Burkert} $\sim$1.2 kpc. 
These findings led us to search for HH objects and collimated outflows in the  BBWo~192E field with narrow-band filters. 

\section{Observations and data reduction}
\label{sec:obs}

The images presented here were obtained on the nights of 13 May 2004
using the NOAO Mosaic II Camera CCD camera at the f/3.1 prime focus
of the 4 meter Blanco telescope at the Cerro Tololo Interamerican
Observatory (CTIO) near La Serena, Chile.   Mosaic II camera is a
8192$\times$8192 pixel array (consisting of eight
2048$\times$4096 pixel CCD chips) with a pixel scale of
0.26$''$ pixel$^{-1}$ and  a field of view 35.4$'$ on a side.

Narrow-band filters centered on
6569\AA\ and 6730\AA\   with a FWHM bandwidth
of 80\AA\  were used to obtain  H$\alpha$ and [S~\textsc{ii}] images.
A Sloan Digital Sky Survey (SDSS) i'  filter centered on
7732\AA\ with a FWHM of 1548\AA\  was used for continuum imaging.
A set of five dithered 600 second exposures were obtained in
H$\alpha$ and [S~\textsc{ii}] using the standard MOSDITHER pattern
to eliminate cosmic rays and the gaps between the
individual chips in Mosaic.  A dithered set of five 180 second
exposures were obtained  in the broad-band SDSS i-band filter to
discriminate between H$\alpha$, [S~\textsc{ii}], and continuum emission.

Images were reduced in the standard manner using IRAF.  Following
bias subtraction, cosmic ray removal, and flat fielding using dome flats,
images were combined using the MSCRED package in IRAF.
Due to insufficient observational time over the whole images there are remaining parts with low S/N.

\section{Results}
\label{sec:res}

\subsection{Visual wavelength imaging}

\begin{figure}
 \centering
  \includegraphics[width=0.45\textwidth]{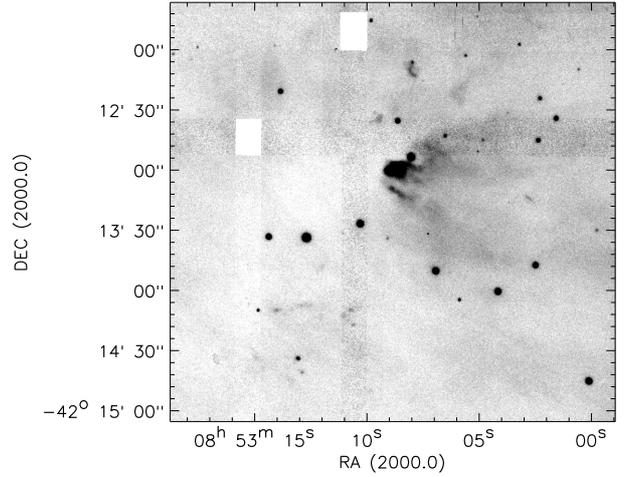}
   \caption{BBWo~192E nebula and new HH objects around it in [S~\textsc{ii}] filter. Rectangular artifacts are produced during the image processing.}
   \label{sii} 
\end{figure}

\begin{figure}
 \centering
  \includegraphics[width=0.45\textwidth]{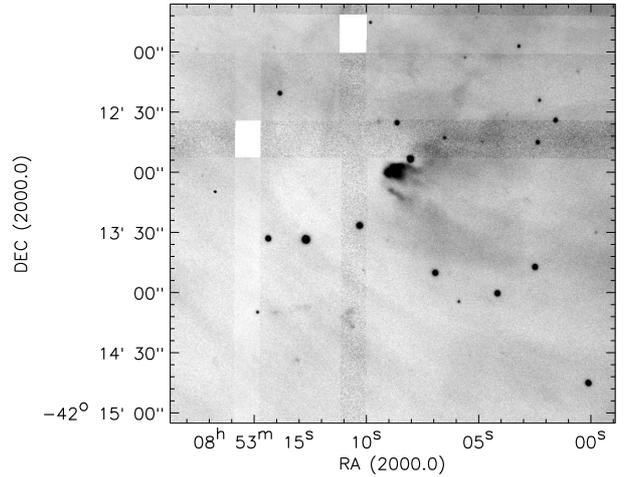}
   \caption{ Same field in H$\alpha$ filter.}
   \label{ha} 
\end{figure}

\begin{figure}
 \centering
  \includegraphics[width=0.45\textwidth]{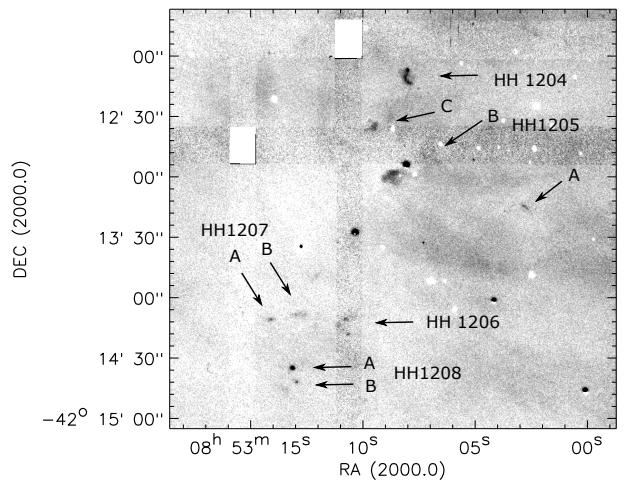}
   \caption{Same field in [S~\textsc{ii}] filter after continuum subtraction. HH objects are marked by arrows. }
   \label{s_i}
\end{figure}

The images reveal several small shock-excited emission nebulae in two or three groups seen in the H-alpha and [S~\textsc{ii}] narrow-band filters but not in the broad i'-band filter.   Given their morphology,  they are likely to be HH objects which are too faint to have been seen in the previous searches.   Their coordinates are given in Table~\ref{HHcoords} and images  are shown on Fig.~\ref{sii} and Fig.~\ref{ha}.    Fig.~\ref{s_i}, where they can be seen better, shows the subtraction of a scaled version of the i'-band image from the [S~\textsc{ii}] image, with the intensity adjusted so that stars have similar counts in the i' and [S~\textsc{ii}] images. The brightest
part of BBWo~192E nebula
also remains visible in this figure, which  may indicate the presence of reflected emission from the central source. We checked the  DECaPS survey \citep{decaps} in which these HH objects are also detected.  However,  they cannot be distinguished by color from the other red stars and nebulous wisps. 
\begin{table}
\caption{List of the new HH objects}
\label{HHcoords}
\centering
\begin{tabular}{l l l}
\hline \hline
Knot & $\alpha $(J2000) & $\delta $(J2000) \\
    & \ \ h \ m \ s & \ \degr\  \arcmin\ \arcsec\ \\
\hline

HH 1204   & 08 53 08.1 &       -42 12 10 \\
HH 1205C & 08 53 02.8 &     -42 13 14 \\
HH 1205B  & 08 58 07.0 & -42 12 49 \\
HH 1205A & 08 53 09.5 &     -42 12 34 \\
\hline
HH 1206 & 08 53 10.9 & -42 14 13 \\
HH 1207A      & 08 53 14.0 & -42 14 10 \\
HH 1207B      & 08 53 12.7 & -42 14 07 \\
HH 1208A  & 08 53 13.1 & -42 14 34 \\
HH 1208B  & 08 53 12.9 & -42 14 41 \\
\hline\hline
\newline
\end{tabular}
\end{table}

As seen in Figs.~\ref{sii} -- \ref{s_i}, all HH knots are brighter in [S~\textsc{ii}] emission. One group of HH objects is located near the northern side of reflection nebula. Two elongated streaks superimposed on the BBWo~192E nebula, labeled as knots A and B in Fig.~\ref{s_i}, along with knot C located north of the reflection nebula core, form the chain we call HH~1205.  The three components, HH~1205 A, B and C, form a nearly straight line and therefore probably trace shocks in one outflow with a full extent of about 85\arcsec,  corresponding to a projected length about a half of parsec at the assumed distance to this complex. To be sure that the streak HH~1205 B is a real emission structure and not a remnant of continuum subtraction, we want to note that it is invisible in H$\alpha$ image and, on the other hand,  the brighter nebulous structure, located to the south from it, is completely subtracted in Fig.~\ref{s_i}.  

HH~1204 is an arcuate feature located about 30\arcsec\ north of the HH~1205 chain, practically visible only in [S~\textsc{ii}].

A second group of HH objects, HH~1206, HH~1207, and HH~1208,  located about 90 to 120\arcsec\ southeast of BBWo~192E. HH~1206,  resembles a bow-shock containing several small knots.  HH~1207 consists of two streaks elongate east-to-west and may trace parts of a collimated flow.    HH~1208 comprises two nearly-stellar  knots.  They are projected on several faint,  nebulous wisps which may be reflection nebulae, given their visibility in the i' image.

Thus, the BBWo~192E field contains at least 9 HH objects tracing several distinct outflows.  However, none are associated with visual wavelength stars  that are likely driving sources.    The stars that drive these outflows are more likely to be highly obscured.   The observed HH objects may trace material ejected from the cloud into the relatively extinction-free foreground.  

\subsection{Infrared data}

\begin{figure}
 \centering
  \includegraphics[width=0.47\textwidth]{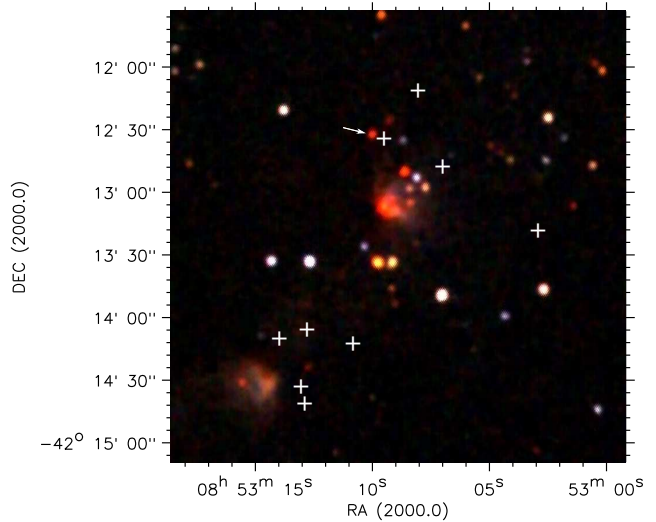}
   \caption{ Color representation of the investigated field from 2MASS survey (\textit{J} - blue, \textit{H} - green, \textit{K} - red). Two infrared nebulae are prominent. The positions of HH objects are marked by white crosses. Star \#28 a.k.a. IRS 26-35 \citep{Burkert,Massi} is pointed by arrow.}
   \label{IR}
\end{figure}

We searched the literature for near-IR studies and inspected several public-domain data sets such as 2MASS and WISE to search for the probable driving sources these HH objects.  \citet{Dutra} found a small infrared cluster (\#22 in their list)\ associated with  BBWo~192E, the existence of which was also suspected in the study of \citet{Burkert}.  This cluster includes also a small nebula, which is the IR analog of BBWo~192E, and is  seen in the 2MASS survey (Fig.~\ref{IR}). Star \#28 from the work of \citet{Burkert}, visible only at IR wavelengths, coincides with the straight line connecting HH~1205 A, B and C.   As is discussed in the same paper, star \#28 is likely to be a pre-main-sequence (PMS) star inside the cloud. It has very red colors according to  \citet[star IRS 26-35]{Massi} and  is a likely candidate source powering the HH~1205 outflow.   There are no infrared sources near HH~1204.  

The near-IR (J, H, K-band) 2MASS images reveal another small, near-IR reflection nebula $\sim$2\arcmin\ to southeast of BBWo~192E in the vicinity of the southeastern group of HH objects
(see Fig.\ref{IR}; J2000.0 coordinates,  RA = $8^{h}\ 53^{m}\ 15^{s}$, Dec =$-$42\degr\ 14\arcmin\ 31\arcsec).   This nebula is not seen in the visual wavelength images  and was not previously mentioned in any catalogs or surveys.     The 2MASS Point Source Catalog lists seven sources inside the nebula, at least some of which may be embedded young stars.  Thus, the existence of a compact infrared cluster of PMS stars inside this nebula seems probable.    HH~1206 and HH~1207 are located northwest of this wedge-shaped IR reflection nebula;  HH~1206 is close to its axis of symmetry. However,  it is unclear which, if any of the IR sources, visible in this region, powers HH~1206, 1207, and 1208.

\begin{figure}
 \centering
  \includegraphics[width=0.45\textwidth]{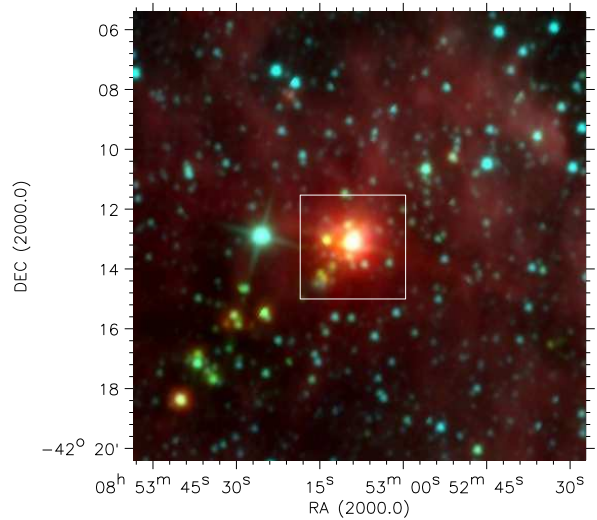}
   \caption{ Color image of a wide field around BBWo~192E from WISE survey (blue - 3.3 $\mu$m, green - 4.7 $\mu$m,  red - 22 $\mu$m).  The area, presented in Figs \ref{sii} -- \ref{IR}, is shown by white square. }
   \label{wise_w}
\end{figure}

\begin{figure}
 \centering
   \includegraphics[width=0.45\textwidth]{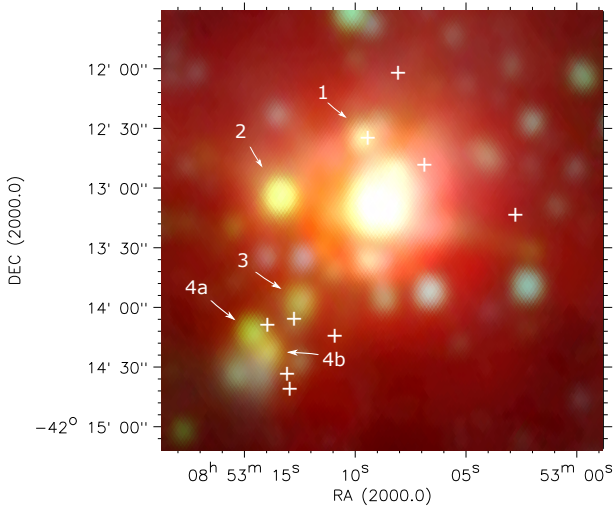}
   \caption{ The  central part of the WISE field shown in Fig.\ref{wise_w}. The bright object in the center is IRAS 08513-4201. The positions of HH objects are marked by white crosses.The sources, described in text, are marked by numbers: 1) - J085309.91-421232.6; 2) - J085313.68$-$421301.6; 3) - J085312.79-421355.2; 4a,b) - J085314.89-421411.4 and J085314.29-421420.2.}
   \label{WISE}
\end{figure}

The clustering of embedded sources around BBWo~192E is apparent in the longer mid-IR WISE survey ($\lambda$ = 3.3, 4.7, 12, and 22 $\mu$m; Wright et al. 2010) (Fig.~\ref{wise_w}).   Wider field-of-view images reveal several additional infrared sources in the region.    IRAS 08513--4201, located near the apex of the BBWo~192E nebula, dominates the field.   Many additional sources not seen in the 2MASS images are also visible.   The most  prominent objects are marked in Fig.~\ref{WISE} and  described below. Their designations are those of the allWISE source catalog along with the numbers from the list of J, H, and K photometry by \citet{Massi}.

\textbf{J085309.91-421232.6 (IRS 26-35)}. This source was  discussed above.
The WISE data confirm its large mid-IR brightness which peaks at 22 $\mu$m (no longer-wavelength photometric data can be found), making it likely to be a young stellar objects(YSO).   It may be the source of the HH~1205 outflow.

\textbf{J085313.68-421301.6 (IRS 26-19)}.
This source, located 48\arcsec\ east from IRAS 08513--4201, is absent at all 2MASS wavelengths but  visible in the WISE 3.3 $\mu$m image.
It becomes  prominent at longer wavelengths, second only to IRAS 08513--4201. It  illuminates the cone-shaped IR reflection nebula, oriented northward.   It is invisible in the  2MASS K-band and in the K-band image in \citet{Burkert}. However, it appears near the limit ($K=16.85$) in the  image presented by \citet{Massi}  which makes it one of the reddest objects in the field. 

\textbf{J085312.79-421355.2 (IRS 26-52)}. This object  located between two IR reflection nebulae described above is absent in the 2MASS catalog, but according to \citet{Massi} is brighter ($K=15.39$) than the previous object. On WISE images it  appears embedded in the faint nebula. However, it is not as red as J085313.68$-$421301.6.

\textbf{J085314.89-421411.4 and J085314.29-421420.2}.
This is a pair of the sources on the north-western side of the newly described IR reflection nebula. They are located outside of the area, studied by \citet{Massi}. The eastern  source is brighter in near IR and can be found in the 2MASS all-sky PSC; the western source  becomes  brighter at 12  $\mu$m in the WISE image.   These  sources may be the main illuminators of the reflection nebula. They are candidate drivers of the outflows containing  HH~1207 and HH~1208. 
 
\subsection{Proper motions and distance}

Most stars in the IR cluster inside BBWo~192E are invisible in the visual wavelength range.   Nevertheless, new astrometric data from the GAIA DR2 catalog allow a search for additional members of this cluster which are not too embedded.

We studied the field around IRAS 08513-4201 within a 3\arcmin\ radius.  The GAIA DR2 catalog contains 61 stars. From these we selected stars with parallaxes in the 0.7--1.8 mas range, taking into account  the systematic correction  +0.045 mas (see, e.g. \citealt{SME}).  The
exact value of this correction is not significant at the distances less
than 1 kpc. 

\begin{figure}
 \centering
  \includegraphics[width=0.45\textwidth]{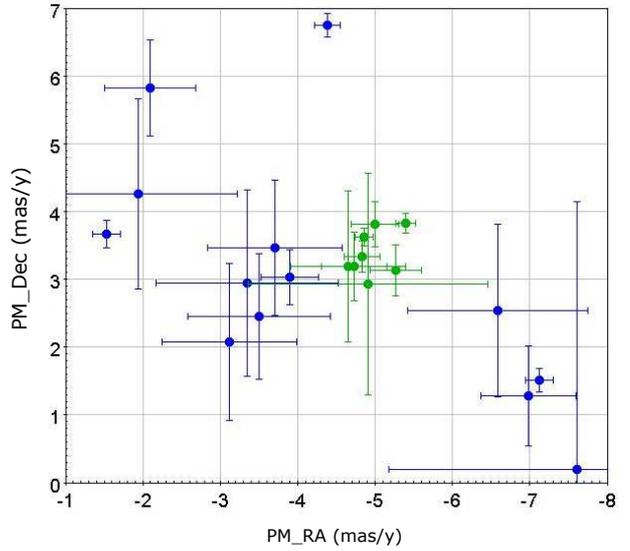}
   \caption{The diagram of proper motions and their error bars for the stars in BBWo~192E field. The stars with most similar PM are marked by green color.  }
   \label{PM}
\end{figure}

\begin{figure}
  \centering
   \includegraphics[width=0.45\textwidth]{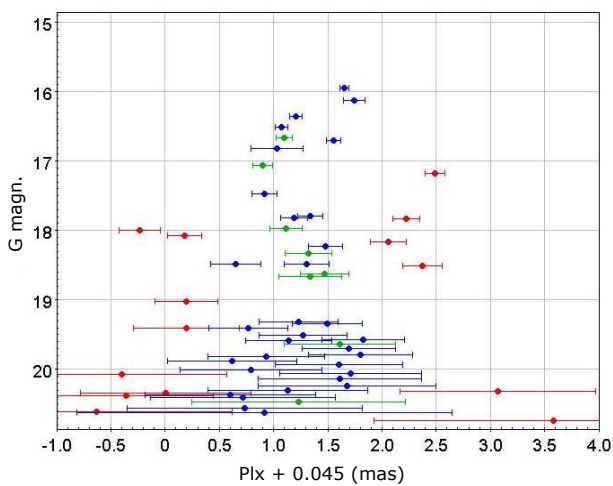}
   \caption{The correlation between visible magnitudes and parallaxes for all stars in BBWo~192E field, measured in GAIA DR2. The stars with parallaxes in 0.7--1.8 mas range are shown by blue, stars with similar PM -- by green. All other stars in the field are marked with red color.  }
   \label{Par}
\end{figure}

Fig.~\ref{PM} shows the distribution of their proper motions (PM) in right ascension and declination.     There are 8 stars with similar proper-motion values.  These stars  have similar parallaxes (see Fig.~\ref{Par}) with a mean value of  $1.32 \pm 0.18$ mas corresponding to a mean distance $800 \pm 100$ pc.    Their mean PM is  $-4.89 \pm 0.19$ mas/y (RA) and $+3.17 \pm 0.28$ mas/y (Dec). One more argument for the possible relationship between this group of stars and the IR cluster is their very red colors (BP$-$RP values are in 2.5--3.7 range).

There are 6 additional faint stars with G magnitude in the 18.5--20.2 range whose PMs  are similar to the 8 brighter stars, but  their parallax and proper-motion errors are large. Thus, these stars may also be members of the cluster. However,   they are located at greater projected distance from the center of the cluster than the eight selected stars. 

\begin{figure}
 \centering
  \includegraphics[width=0.45\textwidth]{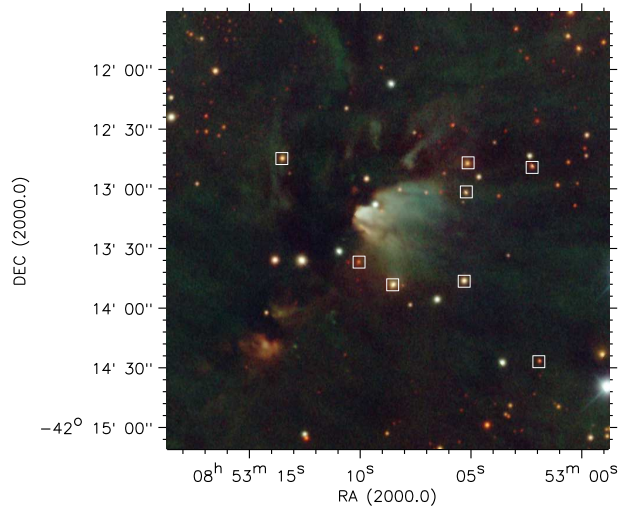}
   \caption{Eight stars with close PM and distances, which are the probable members of BBWo~192E cluster, are marked by white squares on the color image of the field, taken from DECaPS survey.}
   \label{decaps}
\end{figure}

We show the positions of eight probable member stars in Fig.~\ref{decaps}. For completeness, in Table ~\ref{stars} we list their positions and distances according to the catalogue of \citet{BJ}. These distances are estimated by the Bayes method, taking into account  selection effects. The mean distance, computed by these data, is 940 pc.  The difference in values can be explained by various
corrections for the systematic error in parallaxes.

Six of these stars have IR photometry in  \citet{Burkert}, three -- in  \citet{Massi}.   One of these stars is an emission-line star,  ESO-H$\alpha$~259 \citep{PR} .

\begin{table*}
\caption{List of the probable members of BBWo 192E cluster}
\centering
\begin{tabular}{c l l c c c}
\hline \hline
DR2Name & RA (ICRS), degr & Dec (ICRS), degr & & Distance, pc & \\
 & & & Most probable & Min & Max\\
\hline
Gaia DR2 5524333136802415232 & 133.24088919944 & $-$42.24554393145 & 733 &    468 &  1573 \\
Gaia DR2 5524356574444023168 & 133.27889063691 & $-$42.23056943224 &  931 &       869  &     1005 \\
Gaia DR2 5524356608803767680 & 133.26043332165 & $-$42.22976576506 &  1143 &     1030  &    1283 \\
Gaia DR2 5524356707582850176 & 133.30762774432 & $-$42.20570767714 &  924   &    804   &   1084 \\
Gaia DR2 5524356673223197440 & 133.28786696999 & $-$42.22603423908 & 1493   &    674    &  3149 \\
Gaia DR2 5524356845021802240 & 133.25994547813 & $-$42.21249223335  & 791  &     667     &  970 \\
Gaia DR2 5524356845021803392 & 133.25961283028 & $-$42.20670598232 &  804  &     624     & 1124 \\
Gaia DR2 5524356879381541632 & 133.24276963092 & $-$42.20732210284 &  709  &     605   &    855 \\
\hline\hline
\newline
\end{tabular}
 \label{stars}
\end{table*}

\begin{table*}
\caption{Emission and nebulous stars probably related to SL~4 dark cloud}
\centering
\begin{tabular}{c l l c c c}
\hline \hline
Gaia DR2 name & Other names &  Nebula & & Distance, pc & \\
 & & & Most probable & Min & Max\\
\hline
 5524522218442923520 & ESO-H$\alpha$~2348 & BBWo~192C  & 865 &   839 &  893 \\
5524545823586550400 &  & BBWo~192D, GN 08.51.1.01 &  968 &       768  &     1301 \\
5524519989361683840 & ESO-H$\alpha$~249 & BBWo~192B, GN 08.50.5 &  1028 &     1004  &    1053 \\
5524357742675055232 & ESO-H$\alpha$~260 &  &  958   &    876   &   1055 \\
5524520019419933824 & ESO-H$\alpha$~248 &  & 962   &    880    &  1060 \\

\hline\hline
\newline
\end{tabular}
 \label{nebstars}
\end{table*}

Finally, assuming that some of the nebulous stars in the investigated field belong to SL~4 dark cloud, we selected 10 objects from the list of \citet{PR} and BBWo catalog. With the aid of Plx-Gmag and PM diagrams we excluded ESO-H$\alpha$~244 as a foreground object and ESO-H$\alpha$~255, 256 and 257 as background objects. ESO-H$\alpha$~258 has large measurement errors. The remaining 5 stars are listed in Table~\ref{nebstars} with their distances from the catalogue of \citet{BJ}.  Their mean distance is $960 \pm 50$ pc, which confirms their belonging to SL~4 cloud.
However, these nebulae and emission-line stars are slightly more distant than the probable member stars of the BBWo~192E cluster, located in the most opaque part of the cloud.   

GAIA DR2 shows that our distance estimate of 800$\pm$100 pc to the SL~4 cloud is lower than the previously estimated distance \citep{Burkert}.  Our determined distance is similar to the 866 to 965 parsec distance of the Vela~Complex  \citep{Zucker}. For the further estimations of bolometric luminosities we use 800 pc value as based solely on the stellar parallaxes.

\section{Discussion and conclusion}
     
\begin{figure}
 \centering
  \includegraphics[width=0.5\textwidth]{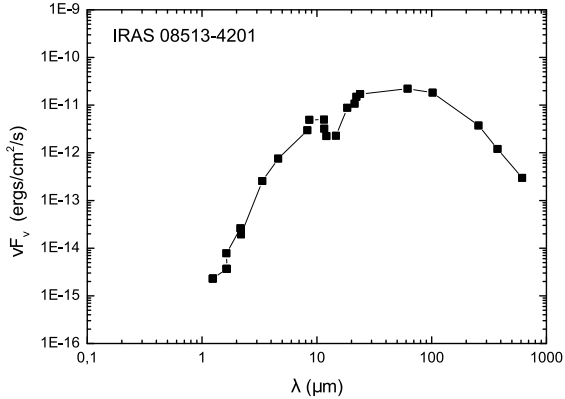}
   \caption{Spectral energy distribution (SED) of IRAS 08513$-$4201, based on the data, selected from literature.}
   \label{IRAS}
\end{figure}

The most luminous infrared source  in the BBW~192E cluster, IRAS 08513$-$4201 (also known as IRS 26-15), was analyzed in \citet{PR} and \citet{Burkert}. It can be unambiguously identified with 2MASS 08530946$-$4213076 and WISE J085309.32$-$421397.3.   The 2MASS point source catalog lists two more nearby sources (08530938$-$4213051 and 08530898$-$4213093), but the close inspection shows that they probably represent the brightest parts of the IR reflection nebula. We added photometry from the MSX6C catalog \citep{MSX}, the AKARI/IRC mid-infrared all-sky survey,   and sub-mm observations using the BLAST telescope \citep{Blast} to built the spectral energy distributions (SED) for this object.   The SED of IRAS 08513$-$4201 is shown in Fig.~\ref{IRAS}.   The SED is  broad, suggesting a wide range of dust temperatures. We estimate bolometric luminosity of IRAS 08513$-$4201 to be L$\approx$1010 $L_{\odot}$,  slightly lower than previous estimates because of the smaller adopted distance.  This objects is likely to be an intermediate-mass HAeBe star.

\begin{figure*}
 \centering
  \includegraphics[width=0.88\textwidth]{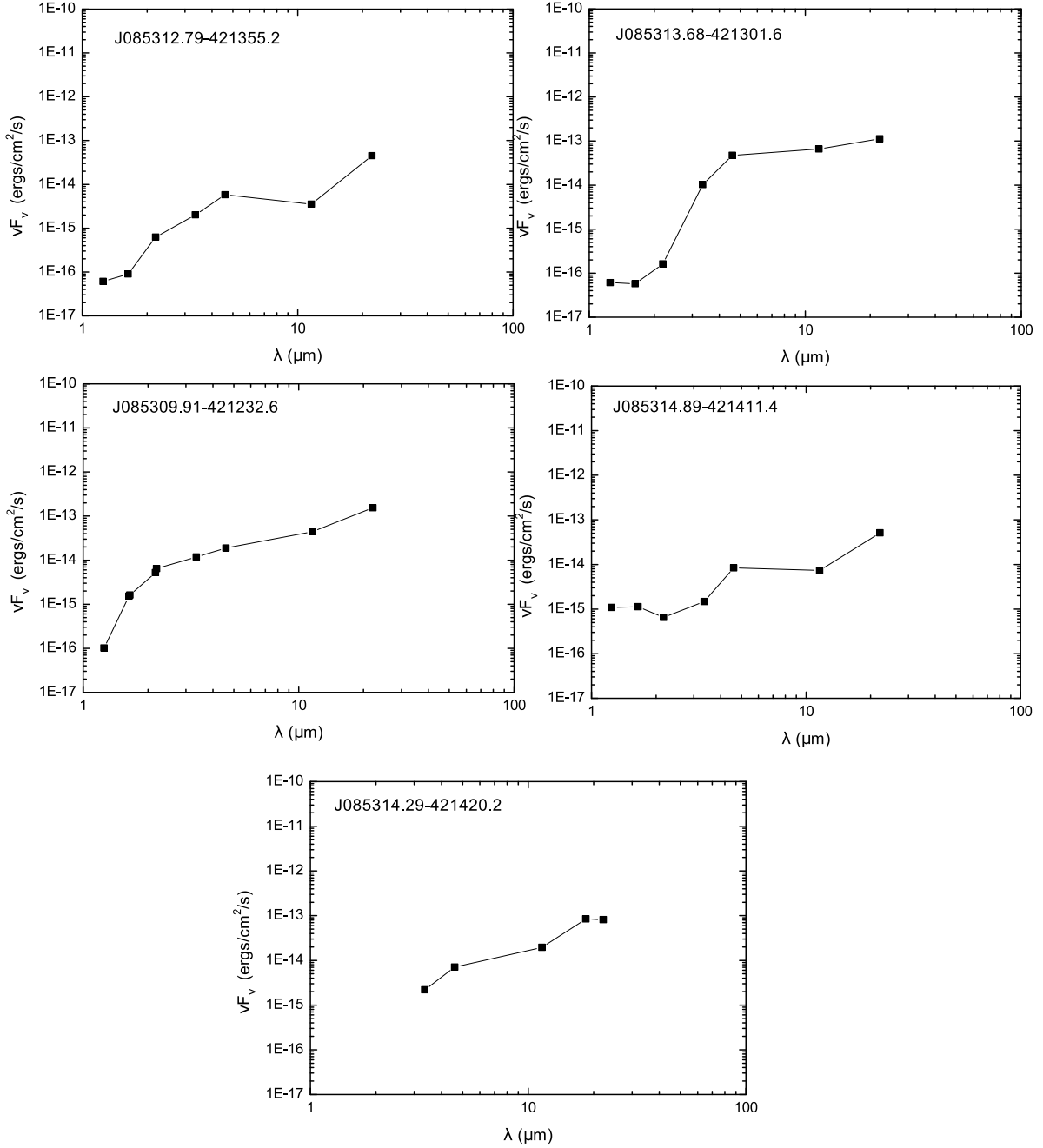}
   \caption{SEDs of other IR sources in the vicinity of IRAS 08513$-$4201.}
   \label{SEDs}
\end{figure*}

Fig.~\ref{SEDs} shows the SEDs of other IR sources, described in Sec. 3.2.
These SED use photometry from 
2MASS and allWISE  since they were not detected in  longer wavelength surveys.    All of these objects emit at wavelengths longer than 2 $\mu$m which demonstrates  high foreground extinction.    We estimated lower-bounds on their luminosities by integrating the SEDs over the observed range. Their luminosities range from from L $>$0.5 to 2.9 $L_{\odot}$.   Since no far-IR data exist, we computed the bolometric corrections using the approach  suggested
by \citet{Cohen}.   The corrected 
bolometric luminosities
of these 5 sources range from 2 to  6 $L_{\odot}$ with J085309.91-421232.6 (IRS 26-35) being the most luminous.   Thus, these are  typical T Tauri class stars.   Thus, the SL~4 cloud contains a small group of forming and young low mass stars surrounding a more massive HAeBe star. 

An interesting feature of the SL~4 cloud is the absence of any  HH flows connected to IRAS 08513$-$4201, the most luminous source. On the other hand, the  existence of the optical HH flows around  both the  reflection nebulae and the cluster of YSOs suggests that star formation in this cloud is on-going.   Assuming a typical flow velocity of 100 km~$s^{-1}$ and a characteristic separation of the HH objects from likely sources, their  kinematic ages  cannot be more than several thousand years. 
The presence of some YSOs and  ionized flows in the visual wavelength images indicates that some members of this group of young stars are not heavily embedded in the dark cloud, and the significant part of line-sight extinction can be produced in the dusty circumstellar disks and envelopes.   

The SL~4 star-forming region is similar to the zone of star formation near CPM~19 \citep[][and references therein]{khanzadyanetal}.   SL~4 may be an interesting target for the study shock-excited H$_{2}$ emission.  Multi-epoch monitoring would be useful for the identification of erratic variables and occasional multi-magnitude accretion-powered flares.  Unfortunately,  this field was not observed by either the Spitzer or Herschel surveys.   If the BBWo 192E nebula is  illuminated by IRAS 08513--4201, then its  visual wavelength or near-IR spectrum  could be obtained by observing the reflection nebula it produces.   Such a study could provide further insights into the mass and evolutionary state of this embedded object.  Finally, given its southern  location,   the BBWo~192E star forming region is ideally placed for deep millimeter and sub-millimeter-wave studies with ALMA.

\section*{acknowledgements}
Authors thank Prof. B.Reipurth for providing the numbers of new HH objects. This work was supported by the RA MES State Committee of Science, in the frames of the research projects number 15T-1C176 and 18T-1C-329.
A.S. Rastorguev also acknowledges RFBR grants Nos. 18-02-00890, 19-02-00611 for partial financial support .This research has made with constant use of ``Aladin sky atlas'' and other products developed at CDS, Strasbourg Observatory, France. This publication makes use of data products from the Two Micron All Sky Survey, which is a joint project of the University of Massachusetts and the Infrared Processing and Analysis Center/California Institute of Technology, funded by the National Aeronautics and Space Administration and the National Science Foundation. This publication makes use of data products from the Wide-field Infrared Survey Explorer, which is a joint project of the University of California, Los Angeles, and the Jet Propulsion Laboratory/California Institute of Technology, funded by the National Aeronautics and Space Administration. This publication has made use of data from the European Space Agency (ESA) mission
{\it Gaia} (\url{https://www.cosmos.esa.int/gaia}), processed by the {\it Gaia}
Data Processing and Analysis Consortium (DPAC,
\url{https://www.cosmos.esa.int/web/gaia/dpac/consortium}). Funding for the DPAC
has been provided by national institutions, in particular the institutions
participating in the {\it Gaia} Multilateral Agreement.

\section*{Data availability}

The data underlying this article will be shared on reasonable request to the corresponding author.

\end{document}